\documentclass[journal]{IEEEcsmag}
\ifCLASSINFOpdf
\else
\fi
\usepackage{algorithmic}
\usepackage{mathrsfs}
\usepackage[nolist]{acronym}
\usepackage{tabularx}
\usepackage{pdfpages}
\usepackage{eurosym}
\usepackage{graphicx}
\usepackage{subcaption}
\usepackage{soul}
\usepackage{cite}
\usepackage{wasysym}
\usepackage{multirow}
\usepackage{float}
\usepackage{ragged2e}
\usepackage{makecell, multirow, tabularx}

\usepackage{siunitx, mhchem}
\usepackage{xcolor} 
\usepackage[ruled, lined, longend, linesnumbered]{algorithm2e}
\usepackage{url}
\usepackage{multicol}
\usepackage{verbatim}
\usepackage{array}

\usepackage{bm}
\jvol{XX}
\jnum{XX}
\paper{8}
\jmonth{Month}
\jname{Publication Name}
\jtitle{Publication Title}
\pubyear{Publication year}

\begin{document}
\sptitle{Feature Article: Collective Intelligence in Drones}
\title{Large-scale Package Deliveries with Unmanned Aerial Vehicles using Collective Learning}

\author{Arun Narayanan}
\affil{School of Energy Systems, LUT University, Finland.}

\author{Evangelos Pournaras}
\affil{School of Computing, University of Leeds, Leeds, UK}

\author{Pedro H. J. Nardelli}
\affil{School of Energy Systems, LUT University, Finland.}

%
%

\markboth{THEME/FEATURE/DEPARTMENT}{THEME/FEATURE/DEPARTMENT}
\begin{abstract}\looseness-1 Unmanned aerial vehicles (UAVs) have significant practical advantages for delivering packages, and many logistics companies have begun deploying UAVs for commercial package deliveries. To deliver packages quickly and cost-effectively, the routes taken by UAVs from depots to customers must be optimized. This route optimization problem, a type of capacitated vehicle routing problem, has recently attracted considerable research interest. However, few papers have dealt with large-scale deliveries, where the number of customers exceed $1000$. We present an innovative, practical package delivery model wherein multiple UAVs deliver multiple packages to customers who are compensated for late deliveries. Further, we propose an innovative methodology that combines a new plan-generation algorithm with a collective-learning heuristic to quickly determine cost-effective paths of UAVs even for large-scale deliveries up to $10000$ customers. Specialized settings are applied to a collective-learning heuristic, the Iterative Economic Planning and Optimized Selections (I-EPOS) in order to coordinate collective actions of the UAVs. To demonstrate our methodology, we applied our highly flexible approach to a depot in Heathrow Airport, London. We show that a coordinated approach, in which the UAVs collectively determine their flight paths, leads to lower operational costs than an uncoordinated approach. Further, the coordinated approach enables large-scale package deliveries.
\end{abstract}

\maketitle


%

\begin{acronym}
  \acro{1G}{first generation of mobile network}
  \acro{1PPS}{1 pulse per second}
  \acro{2G}{second generation of mobile network}
  \acro{3G}{third generation of mobile network}
  \acro{4G}{fourth generation of mobile network}
  \acro{5G}{fifth generation}
  \acro{ARQ}{automatic repeat request}
  \acro{ASIP}{application specific integrated processors}
  \acro{AWGN}{additive white Gaussian noise}
   \acro{BER}{bit error rate}
  \acro{BCH}{Bose-Chaudhuri-Hocquenghem}
  \acro{BRICS}{Brazil-Russia-India-China-South Africa}
  \acro{BS}{base station}
  \acro{CDF}{cumulative density function}
  \acro{CoMP} {cooperative multi-point}
  \acro{CP}{cyclic prefix}
  \acro{CR}{cognitive radio}
  \acro{CS}{cyclic suffix}
  \acro{CSI}{channel state information}
  \acro{CSMA}{carrier sense multiple access}
  \acro{DFT}{discrete Fourier transform}
  \acro{DFT-s-OFDM}{DFT spread OFDM}
  \acro{DSA}{dynamic spectrum access}
  \acro{DVB}{digital video broadcast}
  \acro{DZT}{discrete Zak transform}
  \acro{eMBB}{enhanced mobile broadband}
  \acro{EPC}{evolved packet core}
  \acro{FBMC}{filterbank multicarrier}
  \acro{FDE}{frequency-domain equalization}
  \acro{FDMA}{frequency division multiple access}
  \acro{FD-OQAM-GFDM}{frequency-domain OQAM-GFDM}
  \acro{FEC}{forward error control}
  \acro{F-OFDM}{Filtered Orthogonal Frequency Division Multiplexing}
  \acro{FPGA}{field programmable gate array}
  \acro{FTN}{faster than Nyquist}
  \acro{FT}{Fourier transform}
  \acro{FSC}{frequency-selective channel}
  \acro{GFDM}{generalized frequency division multiplexing}
  \acro{GPS}{global positioning system}
  \acro{GS-GFDM}{guard-symbol GFDM}
  \acro{IARA}{Internet Access for Remote Areas}
  \acro{ICI}{intercarrier interference}
  \acro{IDFT}{Inverse Discrete Fourier Transform}
  \acro{IFI}{inter-frame interference}
  \acro{i.i.d.}{independent and identically distributed}
  \acro{IMS}{IP multimedia subsystem}
  \acro{IoT}{internet of things}
  \acro{IP}{Internet Protocol}
  \acro{ISI}{intersymbol interference}
  \acro{IUI}{inter-user interference}
  \acro{LDPC}{low-density parity check}
  \acro{LLR}{log-likelihood ratio}
  \acro{LMMSE}{linear minimum mean square error}
  \acro{LTE}{Long-Term Evolution}
  \acro{LTE-A}{Long-Term Evolution - Advanced}
  \acro{M2M}{Machine-to-Machine}
  \acro{MA}{multiple access}
  \acro{MAR}{mobile autonomous reporting}
  \acro{MF}{Matched filter}
  \acro{MIMO}{multiple-input multiple-output}
  \acro{MMSE}{minimum mean squared error}
  \acro{MRC}{maximum ratio combiner}
  \acro{MSE}{mean-squared error}
  \acro{MTC}{Machine-Type Communication}
  \acro{NEF}{noise enhancement factor}
  \acro{NFV}{network functions virtualization}
  \acro{OFDM}{orthogonal frequency division multiplexing}
  \acro{OOB}{out-of-band}
  \acro{OOBE}{out-of-band emission}
  \acro{OQAM}{offset quadrature amplitude modulation}
  \acro{PAPR}{peak-to-average power ratio}
  \acro{PDF}{probability density function}
  \acro{PHY}{physical layer}
  \acro{QAM}{quadrature amplitude modulation}
  \acro{PSD}{power spectrum density}
  \acro{QoE}{quality of experience}
  \acro{QoS}{quality of service}
  \acro{RC}{raised cosine}
  \acro{RRC}{root raised cosine}
  \acro{RTT} {round trip time}  
  \acro{SC}{small cell}
  \acro{SC-FDE}{Single Carrier Frequency Domain Equalization}
  \acro{SC-FDMA}{Single Carrier Frequency Domain Multiple Access}
  \acro{SDN}{software-defined network}
  \acro{SDR}{software-defined radio}
  \acro{SDW}{software-defined waveform}
  \acro{SEP}{symbol error probability}
  \acro{SER}{symbol error rate}
  \acro{SIC}{successive interference cancellation}
  \acro{SINR}{signal-to-interference-and-noise ratio }
  \acro{SMS}{Short Message Service}
  \acro{SNR}{signal-to-noise ratio}
  \acro{STC}{space time code}
  \acro{STFT}{short-time Fourier transform}
  \acro{TD-OQAM-GFDM}{time-domain OQAM-GFDM}
  \acro{TTI}{time transmission interval}
  \acro{TR-STC}{Time-Reverse Space Time Coding}
  \acro{TR-STC-GFDMA}{TR-STC Generalized Frequency Division Multiple Access}
  \acro{TVC}{ime-variant channel}
  \acro{UFMC}{universal filtered multi-carrier}
  \acro{UF-OFDM}{Universal Filtered Orthogonal Frequency Multiplexing}
  \acro{UHF}{ultra high frequency}
  \acro{URLL}{Ultra Reliable Low Latency}
  \acro{V2V}{vehicle-to-vehicle}
  \acro{V-OFDM}{Vector OFDM}
  \acro{ZF}{zero-forcing}
  \acro{ZMCSC}{zero-mean circular symmetric complex Gaussian}
  \acro{W-GFDM}{windowed GFDM}
  \acro{WHT}{Walsh-Hadamard Transform}
  \acro{WLAN}{wireless Local Area Network}
  \acro{WLE}{widely linear equalizer}
  \acro{WLP}{wide linear processing}
  \acro{WRAN}{Wireless Regional Area Network}
  \acro{WSN}{wireless sensor networks}
  \acro{ROI}{return on investment}
  \acro{NR}{new radio}
  \acro{SAE}{system architecture evolution}
  \acro{E-UTRAN}{evolved UTRAN}
  \acro{3GPP}{3rd Generation Partnership Project }
  \acro{MME}{mobility management entity}
  \acro{S-GW}{serving gateway}
  \acro{P-GW}{packet-data network gateway}
  \acro{eNodeB}{evolved NodeB}
  \acro{UE}{user equipment}
  \acro{DL}{downlink}
  \acro{UL}{uplink}
  \acro{LSM}{link-to-system mapping}
  \acro{PDSCH}{physical downlink shared channel}
  \acro{TB}{transport block}
  \acro{MCS}{modulation code scheme}
  \acro{ECR}{effective code rate}
  \acro{BLER}{block error rate}
  \acro{CCI}{co-channel interference}
  \acro{OFDMA}{orthogonal frequency-division multiple access}
  \acro{LOS}{line-of-sight}
  \acro{VHF}{very high frequency}
  \acro{pdf}{probability density function}
  \acro{ns-3}{Network simulator 3}
  \acro{Mbps}{mega bits per second}
  \acro{EH}{energy harvesting}
  \acro{SWIPT}{simultaneous wireless information and power transfer}
  \acro{AF}{amplify-and-forward}
  \acro{DF}{decode-and-forward}
  \acro{WIT}{wireless information transfer}
  \acro{WPT}{wireless power transfer}
  \acro{FSFC}{frequency selective fading channel}
  \acro{DC}{direct current}
  \acro{FFT}{fast Fourier transform}
  \acro{RF}{radio frequency}
  \acro{SISO}{single-input single-output}
  \acro{RRC}{root raised cosine}
  \acro{TSR}{time-switching relaying}
  \acro{IFFT}{inverse fast Fourier transform}
  \acro{LIS}{large intelligent surfaces}
  \acro{URLLC}{ultra-reliable low-latency communication}
  \acro{ZMCSCG}{zero mean circularly symmetric complex Gaussian}
  \acro{PPSINR}{post-processing SINR}
  \acro{mMTC}{massive machine-type communication}
  \acro{NR}{New radio}
  \acro{RIS}{reconfigurable intelligent surface} 
  \acro{RAN}{radio access network}
  \acro{i.i.d.}{independent and identically distributed}
  \acro{NOMA}{non-orthogonal multiple access}
  \acro{SDN}{software-defined networks}
  \acro{EMC}{edge-mobile computing}
  \acro{D2D}{device-to-device}
  \acro{SG}{smart grid}
  \acro{MEC}{mobile edge computing}
  \acro{5G-SDVN}{5G-enabled software-defined vehicular networks}
  \acro{MSN}{mobile social network}
  \acro{KPI}{key performance indicator}
  \acro{VNF}{virtual network function}
  \acro{ENPP}{edge node placement problem}
  \acro{MILP}{mixed integer linear programming}
  \acro{V2X}{vehicle-to-everything}
  \acro{PST-ResNet}{deep spatio-temporal residual networks with a permutation operator}
  \acro{SC-DSCS}{subpopulation collaboration based dynamic self-adaption cuckoo search}
  \acro{AI}{artificial intelligence}
  \acro{VANET}{vehicular ad hoc network}
  \acro{MHVA}{multi-hop VANETs-assisted offloading strategy}
  \acro{MINLP}{mixed integer nonlinear programming}
  \acro{JPORA}{joint partial offloading and resource allocation}
  \acro{CF-mMIMO}{cell-free massive MIMO}
  \acro{UAV}{unmanned aerial vehicle}
  \acro{HCP}{heterogeneous computing platform}
  \acro{mGFRA}{massive MIMO based grant-free random access}
  \acro{COP}{concatenated orthogonal preamble}
  \acro{SOP}{single orthogonal preamble}
  \acro{mMIMO}{massive MIMO}
  \acro{SJD}{successive joint decoding}
  \acro{SIC}{successive interference cancellation}
  \acro{GF-NOMA}{grant-free non-orthogonal multiple access}
  \acro{AP}{access point}
  \acro{RIS-SSK}{RIS-space shift keying}
  \acro{RIS-SM}{RIS-spatial modulation}
  \acro{AR}{augmented reality}
  \acro{VR}{virtual reality}
  \acro{FOV}{field of view}
  \acro{PEC}{pervasive edge computing}
  
\acro{IMT-2020}{International Mobile Telecommunications-2020}
\acro{ITU-R}{International Communications Union - Radio-communication Sector}
\acro{eMBB}{Enhanced Mobile BroadBand}
\acro{URLLC}{Ultra-Reliable and Low Latency Communications}
\acro{mMTC}{Massive Machine-Type Communications}
\acro{UHD}{Ultra-High Definition}
\acro{3GPP}{Third Generation Partnership Project}
\acro{5G PPP}{5G Infrastructure Public Private Partnership}
\acro{IEEE}{Institute of Electrical and Electronics Engineering}
\acro{FTP}{File Transfer Protocol}
\acro{METIS}{Mobile and wireless communications Enablers for the Twenty-twenty Information Society}
\acro{IID}{independent and identically distributed}
\acro{BNNs}{Binary Neural Networks}
\acro{ITU}{International Telecommunication Union}
\acro{mmWaves}{millimetre waves}
\acro{UAV}{unmanned aerial vehicle}
\acro{GCS}{ground control station}
\acro{FANET}{flying ad-Hoc network}
\end{acronym}

\section{Introduction}\label{introduction}
\chapteri{U}nmanned aerial vehicles (UAVs, popularly known as \textit{drones}) are rapidly transforming many commercial industries today, and the production of civil UAVs alone is predicted to reach \$11.8 billion globally by 2026 \cite{nouacer_towards_2020}. An important civilian application of UAVs is the delivery of packages to different locations \cite{chung2020optimization,madani2022hybrid}. 
These so-called \textit{delivery drones} have many practical advantages for package deliveries, such as speed, automated delivery, control and tracking, and lower costs. 
As early as November 2016, Domino’s dropped off a pizza order at a customer’s door at 11:19 a.m. in Whangaparaoa, Auckland, New Zealand \cite{dronedelivery2023}. Today, several prominent logistics companies, such as Amazon, UPS, and DHL, have begun deploying delivery drones for commercial package deliveries \cite{dronedelivery2023}.
\begin{figure*}[tb]
\centering
  \includegraphics[scale=0.25]{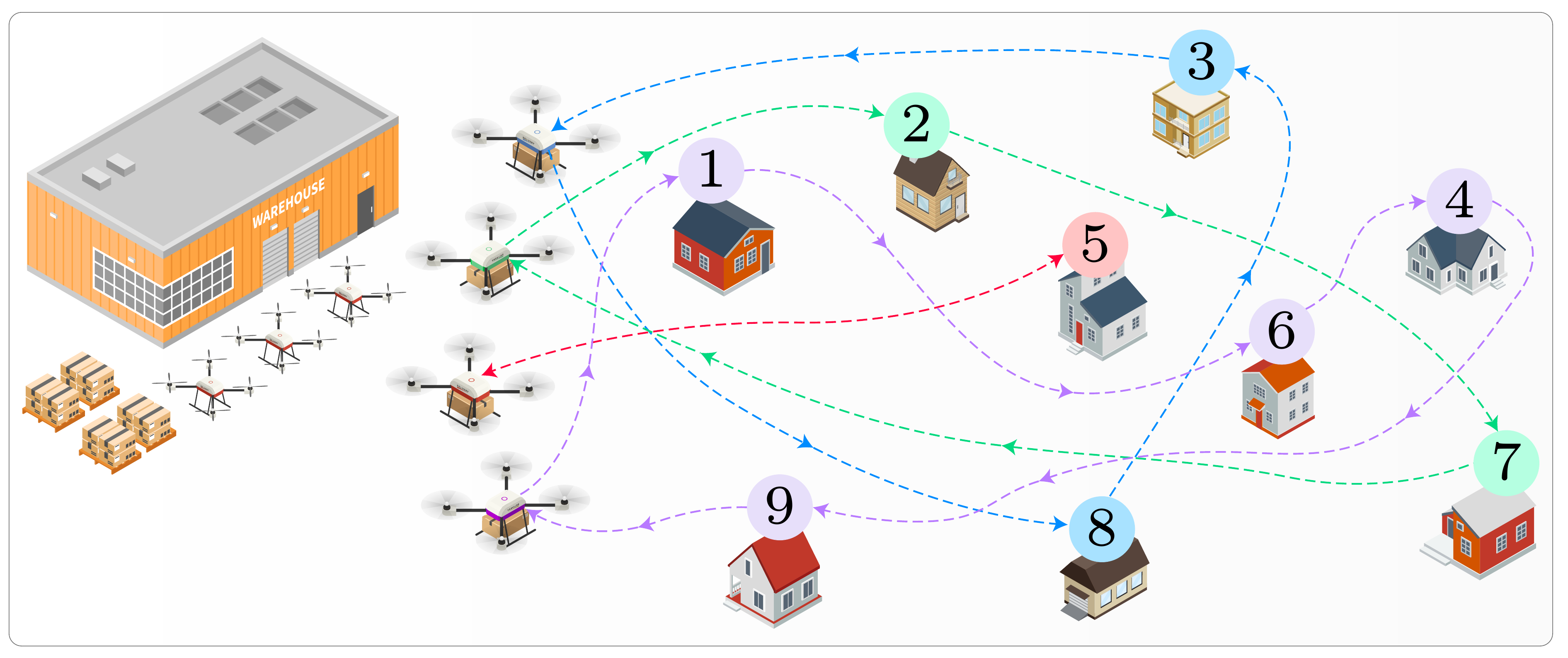}
  \caption{The drone delivery problem: cost-effective routes must be calculated for all the unmanned aerial vehicles (UAVs, or commonly \textit{drones}) that are required to deliver packages from a central depot to $n$ customers. The objective is most often cost, time, or energy minimization, under the constraint that packages have a maximum weight (because of the limited carrying capacity of UAVs). The drone delivery problem is NP-hard and not easily scalable to a large number of customers.}
  \label{fig:drone_del_prob}
\end{figure*}

The traditional, and the most popular, approach to perform drone deliveries, is to pair a UAV with a delivery truck in order to extend the UAV's operations. This model was first introduced by Murray and Chu as the ``flying sidekick traveling salesman problem'' (FSTSP) \cite{murray2015flying}. Here, a delivery truck carries a UAV and all customers' packages from a depot. When the truck is close to a customer's location, the UAV is loaded with the relevant package and launched to make the delivery. The truck then moves to the next customer's location where it rendezvous with the returning UAV. 
Subsequently, researchers extended this basic framework by introducing more complex models that incorporated multiple trucks, multiple UAVs per truck, time windows, heterogeneous UAVs, etc. \cite{chung2020optimization,madani2022hybrid, yin2023branch}.
More recently, a few researchers have considered the so-called ``multivisit problem'' where a UAV carries multiple packages simultaneously and visits multiple costumers \cite{luo_multi-visit_2021, meng2023multi}. 
Most of the studies so far have considered UAV--truck pairing, but some (relatively fewer) studies have also considered swarm-based drone delivery scenarios where numerous drones operate together to make deliveries directly from the depot \cite{chung2020optimization, madani2022hybrid}.

In general, these researches have aimed to optimize the routes of the UAV(s) and truck(s) to achieve a specified objective, typically time or cost minimization. The route optimization problem for drone deliveries is an NP-hard combinatorial-optimization problem that is similar to the classical capacitated vehicle routing problem (CVRP), but with additional complexities and constraints. Hence, while mixed-integer linear programming (MILP) formulations have been used to model the problem scenarios, the predominant approach to solve them has been to use advanced heuristics (e.g., see \cite{murray2020multiple}). However, these heuristics are usually centralized and cannot easily scale for large computational problems, involving large numbers of customers and UAVs. To our knowledge, very few papers have dealt with the problem of large-scale deliveries with UAVs, with the number of customers typically ranging from 10--500 \cite{chung2020optimization, madani2022hybrid}. However, drone delivery problems of significantly larger sizes have important practical applications in future smart cities; according to \cite{arnold2019efficiently}, more than 20,000 package deliveries already need to be carried out daily in a medium-size city such as Brussels. Hence, there is a strong incentive to plan these deliveries as cost-effectively and quickly as possible in delivery businesses at scale.

In this paper, we consider the scenario where multiple UAVs are directly launched from a depot to deliver packages, with each UAV being capable of carrying and delivering multiple packages. Thus, we consider that $u$ UAVs deliver $n$ packages to $n$ customers, starting from a central depot $D$ (Fig. \ref{fig:drone_del_prob}). Such multivist ``drone-only'' scenarios are of increasing practical importance since the range and weight-carrying capacities of UAVs have been increasing in recent times; for example, the Wingcopter 198 can already carry 3 separate packages upto a weight of 5 kg \cite{wingcopterwingcopter1982023}. We also consider that the service operator guarantees that a package will be delivered within a certain time, failing which the customer is compensated (e.g., in the form of a monetary discount). This models real-world scenarios in which delivery companies, such as postal, grocery, and food delivery services, commit to timely shipping and delivery. Thus, our problem is essentially a CVRP with drones and time windows (CVRPDTW) that poses the following question: ``What is a cost-optimal set of routes for a set of UAVs to traverse in order to deliver packages to a given set of customers, if the UAVs have limited carrying capacity and the packages have to be delivered within a certain time window?''

The CVRPDTW deals with the transport of many low-weight packages over a network using several drones in parallel. Hence, we can model this problem as a set of interactions among a network of distributed intelligent software agents, i.e., drones, and their coordinated actions. These interactions lead to an aggregated, or shared, intelligence, often referred to as \textit{collective intelligence} \cite{narayanan2022collective}. 
Collective-intelligence methods involve a coordinated divide-and-conquer approach with high parallelization and scalability, making them suitable for route optimization problems. Collective intelligence can be achieved by \textit{bio-inspired computing methods}, such as ant colony optimization (ACO), or by \textit{collective-learning-based methods}. Bio-inspired computing methods have been employed to solve shortest-path problems such as TSP, VRP, and their variants \cite{yang2013swarm}. 
However, the effectiveness of these methods in obtaining efficient solutions has not been fully tested, especially in the case of nonlinear problems \cite{yang2013swarm}. Moreover, they are not easily scalable to problems of large sizes. For example, ACO needs to store and retrieve the pheromone levels and other data on every edge, leading to significant memory storage; additionally, ants make probabilistic decisions at each city to determine the next visited city, a computationally intensive process \cite{tan2015survey}. 
In this paper, we demonstrate the potential of the alternative approach---c\textit{ollective learning}---to obtain cost-effective solutions to our innovative priority-based drone delivery model, i.e., the CVRPDTW, even when large numbers of customers and UAVs are involved. 

Collective learning is a highly efficient approach for coordinated multi-objective distributed decision-making in multi-agent systems \cite{pournaras_collective_2020}. We develop an innovative methodology that combines a new plan-generation algorithm with collective learning to solve the large-scale CVRPDTW. Two well-known collective-leanring-based optimization heuristic tools in the literature are Combinatorial Optimization Heuristic for Distributed Agents (COHDA) \cite{hinrichs2014cohda} and Iterative Economic Planning and Optimized Selections (I-EPOS) \cite{pournaras_decentralized_2018}. We employ I-EPOS, an unsupervised, collaborative, and highly efficient learning algorithm, since its effectiveness for similar applications in transport and energy have been shown previously \cite{pournaras_collective_2020}. Using our novel plan-generation methodology and specialized EPOS settings, we show for the first time that the drone-delivery problem benefits from a collective distributed optimization approach (as provided by EPOS). We first apply the proposed methodology to an example case considering a depot at Heathrow airport, London. We then scale the problem to 10,000 customers and show that reasonable results could be obtained in around 400 s on a typical desktop PC. We also demonstrate that coordinating the actions of different agents, i.e., UAVs here, leads to a better overall solution than each agent acting independently to pursue its own objective.

\section{Collective Learning} \label{sec:collective_int}
\label{sec:collective_int}

UAVs are often employed to perform certain tasks collectively, e.g., aerial monitoring, tracking, or package deliveries. Here, the UAVs can be managed by humans or they can act independently as autonomous agents that collectively learn, manage, and decide their flight schedules, paths, and patterns to achieve a common goal. Such decision-making through collaboration results in the so-called \textit{collective intelligence}, characterized by information exchange, distributed ``smart'' agents, coordinated decision-making, and adaptive self-management of resources \cite{narayanan2022collective}.

Collective intelligence can be realized by \textit{collective learning}, a form of distributed learning where autonomous agents coordinate their decision-making to collectively learn and manage tasks that can be efficiently performed by coordination, such as reducing electric power peaks \cite{pournaras_decentralized_2018,pournaras_collective_2020}. Collective learning is suitable for combinatorial optimization problems such as the CVRPDTW considered in this paper, where a fleet of autonomous or semi-autonomous UAVs perform a collective and coordinated operation. The UAVs have a set of discrete options, e.g., flight paths and battery power usage, and the collective choices among these options give the overall system performance.

The two main collective-learning algorithms in the literature so far are COHDA and I-EPOS. COHDA was introduced by Hinrichs et al. in 2013 as a decentralized collective-learning heuristic that is applicable to multi-agent systems \cite{hinrichs2013decentralized,hinrichs2014cohda}. COHDA is an iterative asynchronous algorithm that is able to solve multiple-choice combinatorial optimization problems. Subsequently, Pournaras et al. \cite{pournaras_decentralized_2018} proposed a general-purpose decentralized collective-learning algorithm---I-EPOS---in which the agents in a network locally and autonomously self-determine a set of plans, representing their operational flexibility, to meet a goal based on their preferences for the plans. COHDA depends on a system-wide exchange of agents' plans, resulting in a complete exchange of information among the nodes of the network. As a result, COHDA is not easily scalable and there is a significant communication overhead in resource-constraint networks. In contrast, I-EPOS only exchanges local and aggregated plans, making it highly decentralized and privacy-preserving in contrast to related.

The basic idea of the I-EPOS algorithm is as follows (for a more rigorous analysis, see \cite{pournaras_decentralized_2018,pournaras_collective_2020}). Every agent in a network has a finite set of \textit{possible plans}, where a possible plan $p_i$ is a vector $x$ of size $d$ with real values that represent the resource allocation. For example, for a consumer in an electric grid, a plan could be ``$0.45$:$0.39,0.12,0.34,0.85$'' where $0.45$ represents the preference (e.g, cost) for the customer's 4-h energy schedule given by the energy consumed per hour. The consumer would have several such plans, each with different preferences (or costs). From a finite set of possible plans, the agent selects one and only one plan---the \textit{selected plan}---to determine its operation. This agent is connected to several such agents in the network, each with their own selected plans, and an aggregated response is obtained y summing up (element-wise) the selected plans and the cost of the selected plans of every agent. Thus, the selected plans of all agents form a global response vector with an associated global cost. The agents' overall objective is to cooperatively select plans that minimize the global cost. This cooperation is particularly useful when agents' choices depend on each other and as such, the agents minimize a non-linear global cost function.  

\section{Collective Learning for Deliveries with Unmanned Aerial Vehicles \label{sec:problem_and_CI}}
\subsection{Problem model}


Consider a mission in which a depot $D$ has to deliver one package to each of $n$ customers, $C = \{c_1, c_2, c_3,...,c_n\}$, using its fleet of UAVs. Let the set of the weight of the $n$ packages be $W = \{w_{1}, w_{2}, w_{3},...,w_{n}\}$ with $ 0 < w_i \leq J, \; \forall w_i \in W$, where $J$ is the maximum capacity of the UAVs. Thus, each UAV carries at least one package, and the maximum number of required UAVs is $n$. Assume that the depot has a sufficient number of UAVs up to $n$, $U = \{u_1, u_2, u_3,...,u_n\}$, for delivering the packages. A UAV makes one flight per mission in which it delivers multiple packages as long as there is energy left for a return journey to the depot. We then make the following practical assumptions:
\begin{enumerate}
    \item The flight of a UAV begins when it starts from the depot and ends when it returns to the depot.
    \item A UAV performs a maximum of one flight per mission (during which it delivers multiple packages).
    \item A customer is served by at most one UAV per mission.
    \item All customers receive their packages.
    \item A UAV may visit more than one customer per flight.
    \item The UAVs do not visit any non-customer nodes (other than the depot).
    \item All the UAVs remain in constant flight at a constant speed $v_a$ during a mission, except when hovering during parcel delivery.
    
\end{enumerate}

Moreover, we assume that the depot operator charges a price, $p_i$, to a customer $c_i \in C$, such that $p_i$ is proportional to the customer's package weight, $w_i$, and distance to the depot, $d_{Di}$: $p_i = k \times w_i\times d_{Di}$, where $k$ is a constant set by the operator, in $\text{monetary unit}/(\text{grams}\times \text{meter})$. This represents a realistic pricing model of real-world scenarios such as online grocery delivery businesses wherein supermarkets deliver groceries to customers and price their services based on the weight and distance. Further, the operator guarantees that the deliveries will be made within a certain time, failing which a discount is offered to the customer. We assume that if the package is delivered before $t_1$ min, the customer pays the normal price $p$. However, if the package is delivered between $t_1$ and $t_2$ min ($t_2$ > $t_1$), the customer pays $p/2$. And if the delivery happens after $t_2$ min, the customer pays $p/3$. Thus, the discounts are set as $r = \{1, 1/2, 1/3\}$ and a customer $i$ pays $p_i \times r_i$, where $r_i$ depends on the delivery time. 




Since the flight time and distance of a UAV are limited by its weight and the energy stored in its battery, it is important to use an energy consumption model while optimizing deliveries. Several energy consumption models have been proposed in the literature, based on different assumptions \cite{zhang_energy_2021}. In this paper, we use the energy consumption model provided by \cite{dorling2016vehicle} (and modified by \cite{zhang_energy_2021}) as follows:
\begin{align}
E_{pm} = \frac{\left(g\sum_{k=1}^{3}m_k\right)^{3/2}} {v_a\sqrt{2n_r\rho\zeta}} \label{eq.:en_cons}
\end{align}

\noindent Here $E_{pm}$ refers to the energy required for steady UAV flight per unit distance ($J/m$); $g$, acceleration due to gravity ($m/s^2$); \textcolor{red}{$m_k$, mass of UAV component $k$ ($kg$)}; $v_a$, speed of the UAV relative to the air ($m/s$); $n_r$, number of rotors for a rotocopter UAV; $\rho$, air density ($kg/m^3$); and $\zeta$, area of the spinning blade disc of one rotor ($m^2$). 


The CVRPD problem addressed in this paper is then as follows: \textit{determine the number of UAVs $u$ needed and a set of paths for all UAVs $u_i$ ($\forall i = 1,...,u)$, \; $S = \{s_1, s_2, s_3,...,s_u\}$, where each path $s_i$ comprises a set of (visited) customers, such that all $n$ packages are delivered to all $n$ customers with the maximum savings based on the defined assumptions.}

This problem can be set up as a complete graph $G = (V,E)$, where $V = \{0,1,2,…,n\}$ is the set of vertices or nodes (node $0$ is the depot), and $E= \{(i, j):i,j \in V,i\neq j\}$ is the set of edges or arcs. Each customer $i \in V-\{0\}$ has a certain positive demand $w_i\leq J$. A profit or savings $s_{i,j}$ can be associated with each arc $(i, j) \in E$. The graph is a weighted directed graph because the profits depend on the direction of the arc. The objective is to maximize the savings obtained by the system operator for delivering the $n$ packages to the $n$ customers:
\begin{align}
\text{max}\sum_{(i,j)\in E}s_{ij}x^u_{ij} \label{eq:obj}
\end{align}

Here, $x^u_{ij}$ is a binary decision variable $\left(x^u_{ij}\in\{0,1\}\right)$ that decides if a UAV $u$ has taken the route $ij$ (i.e., the edge from $i \rightarrow j$) as follows:
\begin{align}
x^u_{ij}=\begin{cases}
1 & \text{if the arc (i,j) is used}\\
0 & \text{otherwise} \label{eq:dec_var}
\end{cases}
\end{align}

The savings $s_{ij}$ consists of two terms as follows:
\begin{align}
s_{ij} = (p_j \times r_j) - (k \times w_j \times d_{ij})
\end{align}
The first term represents the price paid by the customer at node $j$ to the system operator; here, $r_j$ is the discount based on the time taken to reach $j$, i.e., the total distance traveled to $j$ divided by the speed $v_a$. The second term represents the cost of a UAV to traverse the edge $i$ to $j$, i.e., distance $d_{ij}$.

\begin{algorithm} 
\caption{Plan generation algorithm for drone delivery}
\label{alg:plan-generation} 
\begin{algorithmic}[1]
\STATE \textbf{Inputs}: (1) The set of customer nodes, $C = \{c_1, c_2, c_3,...,c_n\}$; (2) The set of customers' package weights $W = \{w_{1}, w_{2}, w_{3},...,w_{n}\}$; (3) Distances between every node $i$ and $j$, $d_{ij}\; (\forall i,j = 1,....,n$) and distance from the depot $D$ to every node $j$, $d_{Dj}\; (\forall j = 1,....,n$); (4) UAV parameters: weight, battery capacity, battery weight, velocity, number of rotors, area per rotor; (5) air density and acceleration due to gravity; and (6) discount (reward) times $r = {t_1, t_2}$
\STATE Initialize number of UAVs $k = 0$ 
\WHILE{$C \neq \emptyset $}
    \STATE $k = k + 1$, so that current UAV is $u_k$    
	\STATE Randomly select a customer (node) $s_i$ = $c_i, i = 1,...,n$
    \STATE Update set $C$: $C = C \, \setminus \, {c_i}$
    \STATE Initialize plan of UAV $u_k$: $P = \{c_i\}$
    \STATE Initialize feasible set of customers $F = \{\}$ 
    \FOR{j = 1:j++:\,|\,C\,|}
        \IF {{\emph{$w_i + w_j <= J$}} and {\emph{$E_{ij} +  E_{jD} >= E_{rem}$}}}
            \STATE $F = F \cup \{c_j\}$
        \ENDIF
    \ENDFOR
    \STATE Initialize $d = \infty$; $k = 0$
    \FOR{r = 1:r++:\,|\,F\,|}
        \STATE $j =$ index of $r^{th}$ element in $F$
        \IF {$d_{ij} < d$}
            \STATE $k = j$
        \ELSE{}
            \STATE $k = k$
        \ENDIF
    \ENDFOR    
    \STATE Initialize plan of UAV $u_k$: $P = P \cup \{c_j\}$
    \STATE Update set $C$: $C = C \, \setminus \, {c_i}$    
\ENDWHILE
\end{algorithmic}
\end{algorithm}

\subsection{Collective learning approach} \label{subsec:CI_approach}
The CVRPDTW problem set above is an NP-hard problem, and heuristics are required to determine near-optimal, or even reasonable, solutions for large-scale problems. We use one such heuristic, the collective learning algorithm of I-EPOS by combining it with a plan-generation algorithm in the following manner.

We must first generate a finite set of possible plans for every agent in the network, i.e., for every UAV. Each plan corresponds to a set of visited nodes and is associated with the savings accrued by the UAV as a result (Eq. \ref{eq:obj}). For example, one possible plan for a UAV, $u_i$, could be \{1, 3, 5, 9\}, implying that $u_i$ flies as follows: \textit{$\text{depot} \rightarrow \text{node 1} \rightarrow \text{node 3} \rightarrow \text{node 5} \rightarrow \text{node 9} \rightarrow \text{depot}$}. Plan generation is key for achieving optimized solutions, and in this paper, we use the plan-generation algorithm given in Algorithm \ref{alg:plan-generation}. 
Here, first, a UAV is selected and then its plan $P$ is generated. The first visited node, $c_i$, is selected \textit{randomly} from a set of previously unselected nodes (line $5$). Then a set of feasible nodes, $F$, is constructed based on the capacity and energy constraints (line $10$, where $E_{ij}$ and $E_{iD}$ refer to the energy required to go from $i$ to a feasible node $j$ and from $j$ to the depot $D$, respectively). From this feasible set, the \textit{nearest-neighbor criteria} is used to select the next node on the UAV's path, i.e., the node in $F$ that is nearest to $c_i$ is selected. In this manner, a UAV's path is fully constructed. Subsequently, another UAV is selected and its plan is generated similarly. Additional UAVs are selected until all all the customer nodes are selected. Thus, all delivery points are considered ensuring that the approach converges to an optimal solution. Note that the nearest-neighbor approach has also been used indirectly in \cite{arnold2019efficiently}
who obtain solutions for the large-scale CVRP problem. Our approach is supported by their analysis that for many CVRP instances, 95\% of the nodes are connected to around 20 closest neighbors, and 99\% of the customers have two of their nearest 40 nodes as route neighbors.


By selecting only previously unselected nodes and selecting every single node in this process, we ensure that the hard constraints due to the assumptions $1$--$5$ are met in the plan-generation algorithm. Further, the above procedure is repeated $k$ times to generate $k$ plans for each UAV, along with their corresponding savings for each plan. Using a collective-learning approach, I-EPOS combines these plans to determine a cost-effective set of plans with a low global cost (i.e., high savings).

\begin{figure*}[tb]
     \centering
     \begin{subfigure}{0.45\textwidth}
         \centering
         \includegraphics[scale = 0.4]{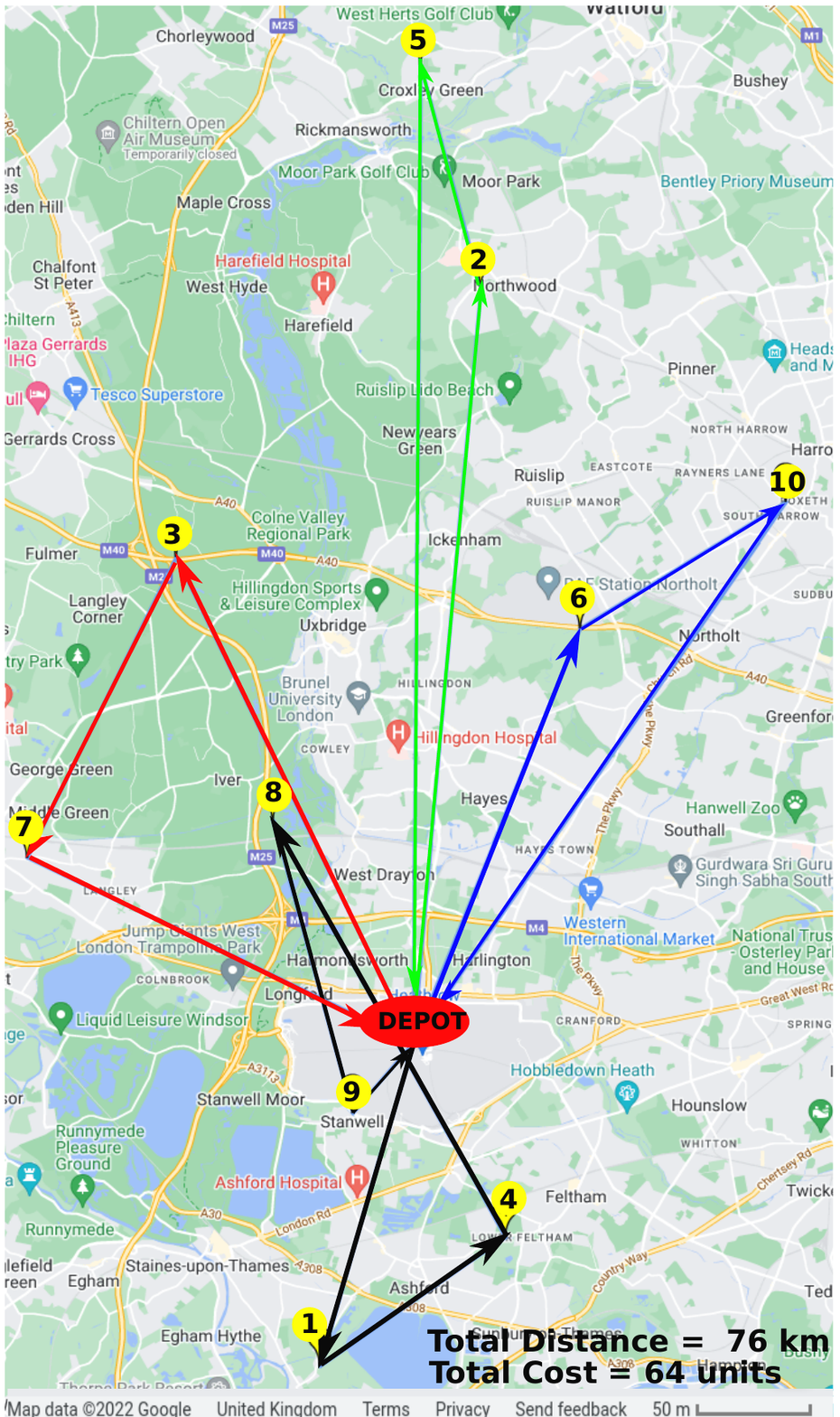}
         \caption{Solution for cost-effective paths \textit{with} coordination among the UAVs.}
         \label{fig:opt_path_coord}
     \end{subfigure}
     \hspace {0.5cm} 
     \begin{subfigure}{0.45\textwidth}
         \centering         
         \includegraphics[scale = 0.4]{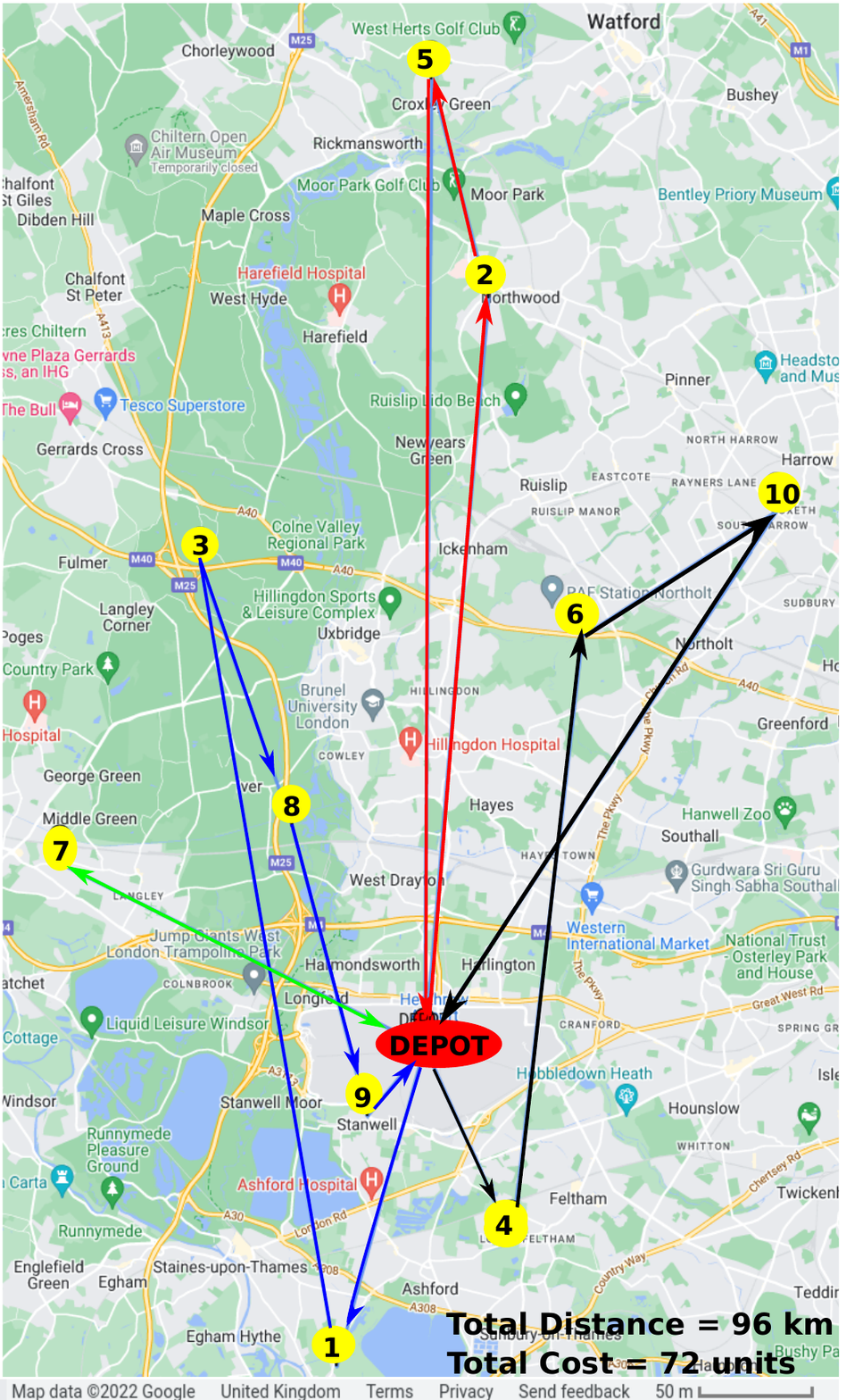}
         \caption{Solution for cost-effective paths \textit{without} any coordination among the UAVs.}
         \label{fig:opt_path_uncoord}
     \end{subfigure}
     \caption{Cost-effective drone paths selected by the I-EPOS algorithm when the selected drones (a) coordinate and (b) do not coordinate. From the depot located at Heathrow Airport, London, $4$ drones deliver packages to $10$ random delivery destinations. In the coordinated case, the UAVs delivered packages to nodes $1, 4, 3,$ and $6$ between $10$ and $20$ min and to the remaining $6$ nodes after $20$ min. In the uncoordinated case, the UAVs delivered packages to node $4$ in less than $10$ min, to nodes $1$ and $7$ between $10$ and $20$ min and to the remaining $6$ nodes after $20$ min.} 
     \label{fig:coord_uncoord}
\end{figure*}

\section{Results and Discussions} \label{sec:results}

\begin{figure*}[tb]
     \centering
     \begin{subfigure}{0.4\textwidth}
         \centering
         \includegraphics[scale=0.55]{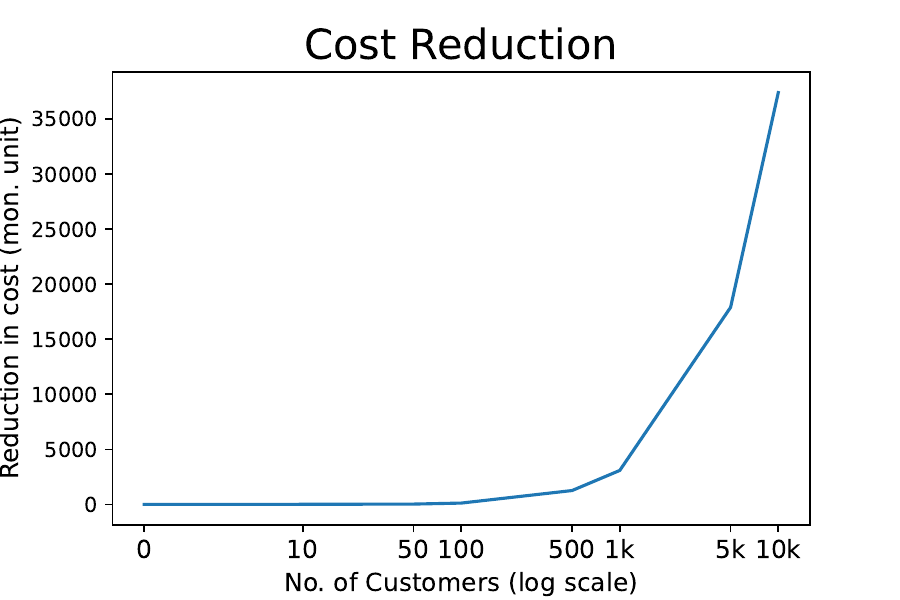}
         \caption{Difference in the savings accrued}
         \label{fig:coord_uncoord_savings}
     \end{subfigure}
     \hspace {2cm} 
     \begin{subfigure}{0.4\textwidth}                  
         \centering
         \includegraphics[scale=0.55]{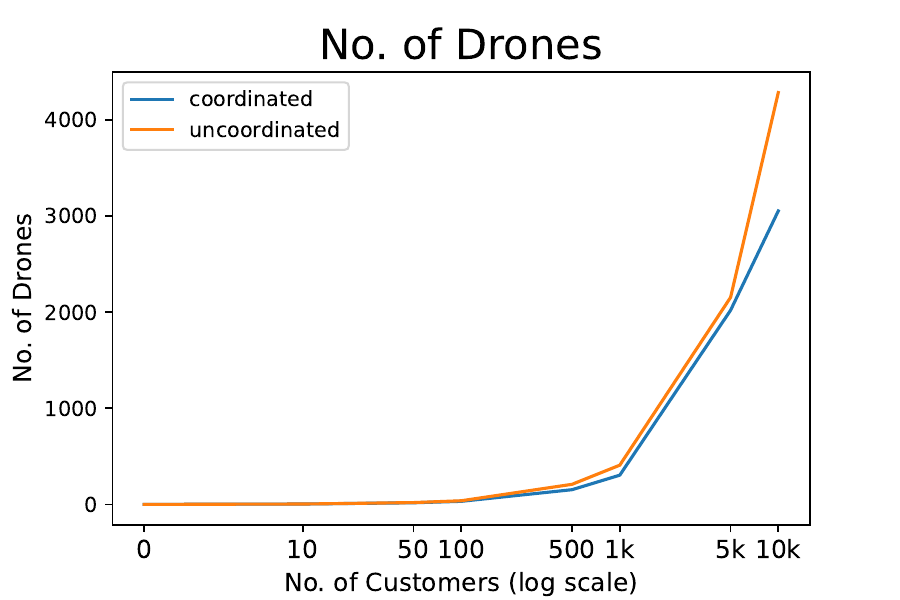}
         \caption{Number of UAVs}
         \label{fig:coord_uncoord_drones}
     \end{subfigure}     
     \caption{Comparison of the cases when the UAVs coordinated among themselves and when they did not coordinate to perform their collective drone delivery mission, as the number of customers was increased. Here, the x-axis represents the number of customers from ($0\text{--}10^4$) plotted in a logarithmic scale.}
     \label{fig:coord_uncoord}
\end{figure*}

We used the following parameters from typical delivery drones that are available in the market. The UAV's frame weight $m_1 = 10$ kg; battery weight $m_2 = 10$ kg; 
capacity $J = 5$ kg; the initial battery capacity = $800$ Wh; 
number of rotors for a rotocopter, $n_r$ = 8; air density at 15 $^{\circ}$C, $\rho = 1.2250$ kg/m\textsuperscript{3}; the area of the spinning blade disc of one rotor $\zeta = 0.27$ m\textsuperscript{2}, and the UAV's velocity $v_a = 10.0$ m/s. Furthermore, we assumed that the depot is located at Heathrow Airport, London, and $k$ was taken as 1 \text{monetary unit}/($\text{grams}\times \text{meter})$. For the ease of presentation in this article, we chose $10$ random delivery destinations in all directions from the depot as the set of customers. We assume that each customer demands a package weighing 0--2 kg, in multiples of 0.5. These weights were estimated randomly as follows: \{1: 0.5, 2: 2.0, 3: 1.0, 4: 0.5, 5: 2.0, 6: 2.0, 7: 2.0, 8: 2.0, 9: 1.0, 10: 2.0\}. Further, the discount (reward) times $r = \{t_1, t_2\}$ were set as $t_1 = 10$ min and $t_2 = 20$ min.  

We generated $10$ plans per agent. The agents interacted for $50$ bottom-up and top-down learning iterations to select their plans. The experiments were repeated $25$ times, with a random placement of the agents in the same balanced binary tree topology. These parameters were chosen after repeated experimentation. Figure \ref{fig:opt_path_coord} shows the paths chosen for the drone deliveries when the UAVs coordinate their path selections. Four UAVs were chosen by the algorithm to carry out the mission. They traveled a total distance of 76 km with a cost of 64 \text{monetary units}, and they chose the following paths: \textit{$0 \rightarrow 1 \rightarrow 4 \rightarrow 8 \rightarrow 9 \rightarrow 0$, $0 \rightarrow 2 \rightarrow 5 \rightarrow 0$, $0 \rightarrow 3 \rightarrow 7 \rightarrow 0$}, and \textit{$0 \rightarrow 6 \rightarrow 10 \rightarrow 0$} (where $0$ refers to the depot). Here, the UAVs delivered packages to nodes $1, 4, 3,$ and $6$ between $10$ and $20$ min and to the remaining $6$ nodes after $20$ min. Similarly, Fig. \ref{fig:opt_path_uncoord} shows the paths when the UAVs do not coordinate among themselves. In this case, the local cost of each UAV is prioritized over the global system cost. Again, $4$ UAVs were chosen but, this time, they traveled a total distance of 96 km with a cost of 72 \text{monetary units}, a 14\% increase, and their paths were \textit{$0 \rightarrow 4 \rightarrow 6 \rightarrow 10 \rightarrow 0$, $ 0 \rightarrow 1 \rightarrow 3 \rightarrow 8 \rightarrow 9 \rightarrow 0$, $0 \rightarrow 2 \rightarrow 5 \rightarrow 0$}, and \textit{$ 0 \rightarrow 7 \rightarrow 0$}. The UAVs delivered packages to node $4$ in less than $10$ min, to nodes $1$ and $7$ between $10$ and $20$ min and to the remaining $6$ nodes after $20$ min.

The proposed method is highly flexible in the sense that non-linear objective functions, e.g., Eq. \ref{eq:obj}, can be handled relatively easily in the plan-generation algorithm that generates the solution subspace. Moreover, importantly, it is computationally feasible to scale up to larger number of customers, even though the solution space increases exponentially with the number of agents ($O(p^a))$, where $p$ is the number of plans per agent $a$, as shown previously in \cite{pournaras_decentralized_2018}. 
Our proposed plan-generation algorithm has a quadratic complexity of $O(n\textsuperscript{2})$. In the case of I-EPOS, the computational complexity is linear to the number of plans and dependent on the number of iterations and the number of children \cite{pournaras_decentralized_2018}. As the number of plans increases, the global cost and convergence speed decrease but the computation time increases (see \cite{pournaras_decentralized_2018} for details). For one simulation round ($50$ iterations), a reasonable solution could be obtained in around $45$ s for 10,000 customers and $10$ plans (in a laptop with Intel\ Core(TM) i5-8400H CPU @ 2.50GHz and 32 GB RAM). On the other hand, for 10,000 customers and $100$ plans, I-EPOS required around 400 s to converge but with a lower global cost. Therefore, to reduce costs, the number of plans can be increased, for example, to $10\%$ of the number of customers, while noting that this increases the time required for convergence. Thus, there is a tradeoff between the simulation time versus the global cost.

Furthermore, scaling to larger number of customers requires additional computations because of the hard constraints that every customer must receive their package and they should be visited by one and only one UAV. When the number of customers is small, as in the example above, all the nodes are selected in both the coordinated and uncoordinated cases. However, when the number of customers is large (>100), we find that in the coordinated case, more than $90\%$ of the nodes are always selected, whereas in the uncoordinated case, only $65\%$ are typically selected (even with 1 iteration). As a result, for large customers, we calculated the results by running the proposed method (plan generation with I-EPOS) again with the remaining unselected customers, until every customer node was selected. Figure \ref{fig:coord_uncoord} compares the cases when the UAVs were coordinated and when they were not coordinated, as the number of customers were increased. As shown in Figure \ref{fig:coord_uncoord_savings}, the cost reduction as a result of coordination increases dramatically with the number of customers. Moreover, the uncoordinated case selects a higher number of UAVs as the number of customers increases, with the difference with the coordinated case showing a clear increasing trend (Figure \ref{fig:coord_uncoord_drones}). Thus, the coordinated approach outperforms the uncoordinated approach, not only increasing the savings but also making it much easier to scale to a large number of customers.

\section{Limitations and Future Work} \label{sec:limitations}
Our proposed methodology has a few limitations that must be overcome before practical deployments. We do not explicitly consider emergencies that can cause real-time flight changes, e.g., battery failures, weather conditions, collisions etc. We also do not discuss the impact of governmental regulations, such as restrictions on UAVs' heights, flight zones, and flying times. Nevertheless, this is a first step toward developing a fully practical system for large-scale drone deliveries. In the future, we will focus on real-time flight coordination to address emergencies, and explore the possibilities of multiple depots, battery recharging, and battery swapping. Another interesting idea is to couple the package-delivery service with other remunerative services, for example, temporarily providing  WiFi coverage to an area, without sacrificing delivery time or efficiency. 

\bibliographystyle{IEEEtran}
\bibliography{ref}

\end{document}